\documentclass[twocolumn]{aastex631}
\usepackage{amsmath}
\providecommand{\sorthelp}[1]{}

\begin{document}

\title{The Optical Extinction Law Depends on Magnetic Field Orientation: The $R_V$--$\psi$ Relation}
\shorttitle{The $R_V$--$\psi$ Relation}

\author[0000-0001-7449-4638]{Brandon S. Hensley}
\affiliation{Jet Propulsion Laboratory, California Institute of Technology, 4800 Oak Grove Drive, Pasadena, CA 91109, USA}

\date{\today}

\begin{abstract}
For aspherical interstellar dust grains aligned with their short axes preferentially parallel to the local magnetic field, the amount of extinction per grain is larger when the magnetic field is along the line of sight and smaller when in the plane of the sky. To the extent that optical extinction arises from both aligned and unaligned grain populations with different extinction properties, changes in the magnetic field orientation induces changes in its wavelength dependence, parameterized by $R_V \equiv A_V/E(B-V)$. We demonstrate that the measured total and polarized extinction curves of the diffuse Galactic interstellar medium imply $R_V$ varies from 3.21 when the magnetic field is along the line of sight ($\psi = 0$) to $R_V = 3.05$ when in the plane of the sky ($\psi = 90^\circ$). This effect could therefore account for much of the large-scale $R_V$ variation observed across the sky ($\sigma(R_V) \simeq 0.2$), particularly at high Galactic latitudes.\footnote{\textcircled{c} 2024. California Institute of Technology. Government sponsorship acknowledged}
\end{abstract}

\section{Introduction}
It has long been recognized that absorption and scattering by interstellar dust particles obscures our view of the distant Universe \citep{Trumpler:1930}. Dust not only dims objects but also reddens their color, a consequence of the fact that dust extinction generally increases with decreasing wavelength in the optical. Accurate determinations of dust extinction are required for everything from construction of 3D maps of the Milky Way to precision cosmology (see, e.g., \citet{Zhang:2024} and \citet{Karim:2024} for recent examples, respectively).

Dust extinction arises from different populations of interstellar grains. Optical extinction is typically attributed to submicron grains, though models disagree on whether these are a single homogeneous population \citep{Hensley:2023} or multiple distinct compositions \citep{Guillet:2018, Siebenmorgen:2023, Ysard:2024}. It is clear from albedo measurements \citep{Lillie:1976}, however, that the prominent 2175\,\AA\ feature that dominates the UV extinction arises from much smaller particles. Variations in the shape of the extinction curve are generally explained by changes in the relative abundances of these species or changes in their size distributions \citep[e.g.,][]{Weingartner:2001}.

Aspherical grains align with their short axes preferentially parallel to the local magnetic field \citep{Purcell:1979}. Since grains absorb and scatter light more efficiently when the applied $E$-field is along their long axis, starlight absorbed by intervening dust is polarized parallel to the magnetic field lines. Robust polarization of starlight at optical wavelengths \citep{Hall:1949, Hiltner:1949} and polarized emission from dust at infrared wavelengths \citep{planck2014-XIX} attest the presence of aligned grains in the interstellar medium (ISM). However, not all grains are aligned. While dust extinction is particularly strong at UV wavelengths, it is only weakly polarized \citep{Martin:1999}. Further, the 2175\,\AA\ feature is nearly absent in polarization \citep{Clayton:1992, Wolff:1997}. These observations point to the fact that small interstellar grains are not well-aligned \citep{Kim:1995}.

Viewed from different angles, an aspherical grain has different extinction cross sections: if aligned with its short axis parallel to the local magnetic field, the grain has a larger extinction cross section when the magnetic field is along the line of sight than when in the plane of the sky \citep{Greenberg:1960}. In general, the wavelength dependence of extinction from aspehrical grains changes with viewing angle, particularly for wavelengths comparable to the grain size \citep{Greenberg:1960b, Greenberg:1987}. For a single population of ice-coated astrosilicate grains under the assumption of suprathermal spinning alignment, \citet{Voshchinnikov:1989} found that $R_V$ changes from 3.38 to 2.97 as the angle between the magnetic field and line of sight varies from 30 to 90$^\circ$. As demonstrated in the present study, dependence of the extinction curve on viewing angle can also arise if both aligned and unaligned grain populations with different extinction properties contribute to the observed extinction.

In this work, we develop a framework for quantifying the amplitude of $R_V$ variations independent of a particular grain model. Relying solely on the observationally determined total and polarized extinction curves of the Milky Way, we demonstrate that $R_V$ changes by 0.16 over the range of viewing angles. This is comparable to the observed variation in $R_V$ across the Galactic plane \citep{Schlafly:2016}.

This paper is organized as follows: Section~\ref{sec:theory} presents the theoretical basis of the $R_V$--$\psi$ relation; Section~\ref{sec:rv_psi} quantifies the amplitude of $R_V$ variation as a function of $\psi$ and discusses implications; Section~\ref{sec:pol} discusses connections between $R_V$ and observations of polarized dust emission; and Section~\ref{sec:summary} provides a brief summary.

\section{Theory} \label{sec:theory}
An aspherical, rotating interstellar grain undergoes several distinct alignment processes. For sake of generality, we consider a triaxial grain with principal axes $a_1 \leq a_2 \leq a_3$ and define $\hat{\bf a}_1$, $\hat{\bf a}_2$, and $\hat{\bf a}_3$ as unit vectors along the respective axes. First, internal alignment aligns the angular momentum vector ${\bf J}$ with the short axis, i.e., ${\bf J} \parallel \hat{\bf a}_1$. Second, rotation induces a magnetic moment in the grain---the Barnett effect---causing ${\bf J}$ to precess around the local magnetic field ${\bf B}$ \citep{Dolginov:1976}. Finally, ${\bf J}$ is brought into alignment with ${\bf B}$ through a combination of radiative torques and (super)paramagnetic dissipation \citep{Draine:1997, Hoang:2016}. 

As long as the grain is spinning suprathermally such that thermal fluctuations cannot randomize the orientation of ${\bf J}$, we have that $\hat{\bf a}_1 \parallel {\bf J} \parallel {\bf B}$ \citep{Purcell:1975, Purcell:1979}. Large submicron grains can be driven to suprathermal rotation by radiative torques, whereas smaller grains cannot be spun up by UV or optical photons \citep{Draine:1996}. Thus, the alignment of interstellar grains depends upon size.

To analyze the extinction properties of a population of interstellar grains of various sizes, we invoke the ``modified picket fence approximation'' \citep[MPFA,][]{Draine:2021}. In this approximation, the extinction cross section of an arbitrarily oriented grain can be inferred from its extinction cross sections in which the photon electric field ${\bf E}$ and propagation direction $\hat{\bf k}$ are oriented along principal axes of the grain. 

We make three additional approximations to simplify the analysis. First, we assume perfect internal alignment, i.e., that all grains rotate about their short axis. Given the rapid timescale for internal alignment \citep{Purcell:1979}, this should be an accurate approximation. Second, we assume that magnetic dipole absorption is small and thus that, in the MPFA, the extinction cross section only depends on which of the principal axes ${\bf E}$ is oriented along. We denote the extinction cross sections with ${\bf E} \parallel \hat{\bf a}_1$, ${\bf E} \parallel \hat{\bf a}_2$, and ${\bf E} \parallel \hat{\bf a}_3$ as $C_{\rm ext}^1$, $C_{\rm ext}^2$, and $C_{\rm ext}^3$, respectively. Since the grain is rotating about $\hat{\bf a}_1$, the extinction cross section of a grain of any orientation can be written in terms of only $C_{\rm ext}^1$ and $(C_{\rm ext}^2+C_{\rm ext}^3)/2$ under these assumptions. Finally, we assume that grains are either perfectly aligned or are else randomly oriented. The fraction of grains that are perfectly aligned $f$ is numerically equivalent to the ``reduction factor'' that arises when considering partial alignment of ${\bf J}$ with ${\bf B}$ \citep{Greenberg:1968, Hensley:2019, Draine:2021}.

While this formulation of extinction from partially aligned grains is inexact, the MPFA has been demonstrated to give results accurate to $\lesssim10\%$ for optical extinction and polarization and is even more accurate when averaging over a distribution of grain sizes \citep{Draine:2021}. We do not anticipate the conclusions of this work to change by more precise averaging over grain orientations.

The following derivation draws closely on \citet{Hensley:2019}, who analyzed the effect of $\psi$ on emission from a population of partially aligned grains \citep[though see also][]{Lee:1985, Draine:2021}. The expressions here for extinction cross sections for grains of a single size and composition are identical to those for the absorption cross sections in \citet{Hensley:2019}. The analyses diverge in the need for the present calculation to account for variation as a function of grain size and composition.

A randomly oriented grain has an extinction cross section

\begin{equation} \label{eq:cext_ran}
    C_{\rm ext}^{\rm ran} = \frac{1}{3}\left(C_{\rm ext}^1 + C_{\rm ext}^2 + C_{\rm ext}^3\right)
    \,.
\end{equation}
Under the assumptions above, if $\psi$ is the angle between the line of sight and ${\bf B}$, a perfectly aligned grain has extinction cross section

\begin{align}
    C_{\rm ext}^{\rm align} &= \frac{1}{2}\left(C_{\rm ext}^2+C_{\rm ext}^3\right)\cos^2\psi \nonumber \\
    &+ \frac{1}{2}\left[C_{\rm ext}^1 + \frac{1}{2}\left(C_{\rm ext}^2+C_{\rm ext}^3\right)\right]\sin^2\psi\,.
\end{align}
Thus the total extinction cross section is

\begin{equation}
    C_{\rm ext}^{\rm tot} = f C_{\rm ext}^{\rm align} + \left(1-f\right) C_{\rm ext}^{\rm ran}\,.
\end{equation}
It is useful to define the cross section 

\begin{equation} \label{eq:cext_pol}
    C_{\rm ext}^{\rm pol} \equiv \frac{1}{2}\left[\frac{1}{2}\left(C_{\rm ext}^2+C_{\rm ext}^3\right) - C_{\rm ext}^1\right]
\end{equation}
so that

\begin{equation}
    C_{\rm ext}^{\rm tot} = \left[C_{\rm ext}^{\rm ran} + f C_{\rm ext}^{\rm pol} \left(\frac{2}{3} - \sin^2\psi\right)\right]\,.
\end{equation}
Summing over grain compositions $i$, the optical depth at wavelength $\lambda$ for a population of grains is

\begin{equation} \label{eq:tau}
    \tau\left(\lambda, \psi\right) = \sum_{i} \int da \left(\frac{dN_i}{da}\right) C_{\rm ext}^{\rm tot}\left(\lambda, a, i, \psi\right)\,,
\end{equation}
where $\left(dN_i/da\right)da$ is the number of grains per cm$^2$ of composition $i$ with size between $a$ and $a+da$. The dependencies on wavelength $\lambda$, composition $i$, grain size $a$\footnote{It is typical to take $a = (a_1a_2a_3)^{1/3}$, but the presentation here does not depend upon a precise specification of $a$.}, and $\psi$ are now indicated explicitly.

It is evident from Equation~\eqref{eq:tau} that the extinction per unit grain mass depends on $\psi$ and that the magnitude of this effect depends on $f$. If $f=0$, all grains are randomly oriented and $\tau$ is independent of $\psi$. If $f=1$, all grains are aligned and the disparity in $\tau$ for $\psi = 0$ versus $\psi = 90^\circ$ is maximal.

Now imagine that there are two populations of dust, one for which $f$ is small and one for which $f$ is close to unity. The relative contributions of these two populations to $\tau$ changes with $\psi$: the amount of extinction from the former varies little with $\psi$ while the amount of extinction from the latter decreases as $\psi$ increases from 0 to $90^\circ$. If these populations differ in the wavelength dependence of their extinction, then the wavelength dependence of $\tau$ must likewise vary with $\psi$.

The polarized extinction arising from the same population of grains producing the total extinction $\tau\left(\lambda, \psi\right)$ is

\begin{equation} \label{eq:extpol}
    p\left(\lambda, \psi\right) = \sum_{i} \int da \left(\frac{dN_i}{da}\right) f\left(a, i\right) C_{\rm ext}^{\rm pol}\left(\lambda, a, i\right)\sin^2\psi\,.
\end{equation}
If we define

\begin{equation}
    \tau^{\rm ran}\left(\lambda\right) \equiv \sum_{i} \int da \left(\frac{dN_i}{da}\right) C_{\rm ext}^{\rm ran}\left(\lambda, a, i\right)\,,
\end{equation}
we can write

\begin{equation} \label{eq:tau_p}
    \tau\left(\lambda, \psi\right) = \tau^{\rm ran}\left(\lambda\right) + p\left(\lambda, 90^\circ\right)\left(\frac{2}{3} - \sin^2\psi\right)\,.
\end{equation}
Given that the polarized extinction $p\left(\lambda\right)$ is observed to have a markedly different wavelength dependence than the mean total extinction $\tau^{\rm ran}\left(\lambda\right)$, Milky Way dust evidently comprises such differing populations. 

\section{The $R_V$--$\psi$ Relation} \label{sec:rv_psi}

\begin{figure}
     \centering
     \includegraphics[width=\columnwidth]{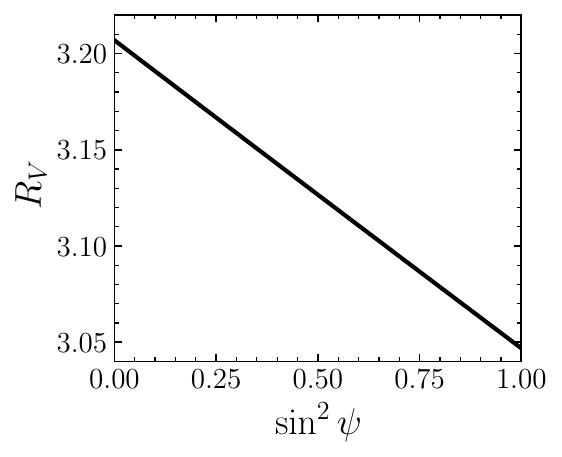}
     \caption{Variation of $R_V$ with the angle $\psi$ between the interstellar magnetic field and the line of sight. The relation is fully prescribed by observationally determined quantities as per Equation~\eqref{eq:rv_psi}.} \label{fig:rv}
 \end{figure}

 \begin{figure}
     \centering
     \includegraphics[width=\columnwidth]{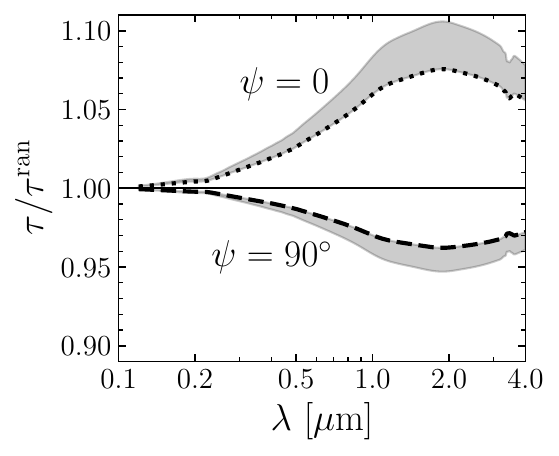}
     \caption{Extinction curve for $\psi = 0$ (dotted) and $\psi = 90^\circ$ (dashed) relative to the orientation-averaged extinction curve ($\sin^2\psi = 2/3$, solid). The curves are derived from Equation~\eqref{eq:tau_p} using $\tau^{\rm ran}\left(\lambda\right)$ and $p\left(\lambda, 90^\circ\right)$ of the diffuse Galactic ISM as compiled by \citet{Hensley:2021}. The shaded region corresponds to $\left(p_V/E(B-V)\right)_{\rm max}$ between 0.13 and 0.182\,mag$^{-1}$.} \label{fig:ext}
 \end{figure}

The optical extinction law is often parameterized by $R_V \equiv A_V/E(B-V)$, i.e., the ratio of the extinction in the Johnson $V$ band $A_V$ to the reddening $E(B-V) \equiv A_V - A_B$, where $A_B$ is the extinction in the Johnson $B$ band. Remarkably, Equation~\eqref{eq:tau_p} allows us to quantify the impact of $\psi$ on $R_V$ in a model-independent way. If many measurements of $\tau\left(\lambda, \psi\right)$ are combined to make a composite extinction curve, variations in $\psi$ are averaged out, yielding a determination of $\tau^{\rm ran}\left(\lambda\right)$. The wavelength dependence of the polarized extinction is independent of $\psi$ (Equation~\eqref{eq:extpol}) and has been well characterized. $p\left(\lambda, 90^\circ\right)$ can be determined observationally by identifying the lines of sight with the most polarization per unit extinction.

Let $R_V^{\rm ran}$ be the $R_V$ value measured on a composite Milky Way extinction curve, i.e., $R_V^{\rm ran} = \tau^{\rm ran}_V/\left(\tau^{\rm ran}_B - \tau^{\rm ran}_V\right)$, and let $\left(p_V/\tau_V\right)_{\rm max}$ denote $p_V\left(\psi=90^\circ\right)/\tau^{\rm ran}_V$. Then $R_V\left(\psi\right)$ can be derived from Equation~\eqref{eq:tau_p} as

\begin{equation} \label{eq:rv_psi}
    R_V\left(\psi\right) = R_V^{\rm ran} \cdot \frac{1 + \left(\frac{p_V}{\tau_V}\right)_{\rm max}\left(\frac{2}{3} - \sin^2\psi\right)}{1 + R_V^{\rm ran} \left(\frac{p_V}{\tau_V}\right)_{\rm max}\left(\frac{p_B}{p_V} - 1\right) \left(\frac{2}{3} - \sin^2\psi\right)}\,.
\end{equation}

The mean Galactic extinction curve has $R_V^{\rm ran} = 3.1$ \citep{Fitzpatrick:2019}. At optical wavelengths, the polarized extinction curve follows the ``Serkowski Law'' \citep{Serkowski:1975}

\begin{equation}
    p\left(\lambda\right) \propto {\rm exp}\left[-K \ln^2\left(\lambda_{\rm max}/\lambda\right)\right]\,,
\end{equation}
where $K = 0.87$ and $\lambda_{\rm max} = 5500$\,\AA\ are typical parameters \citep{Whittet:2003}. Approximating $p_B \approx p\left(4400\,\text{\AA}\right)$ and $p_V \approx p\left(5500\,\text{\AA}\right)$, $p_B/p_V = 0.96$. The maximum value of $p_V/E(B-V)$, corresponding to when $\psi = 90^\circ$, has been determined to be at least 13\%\,mag$^{-1}$ \citep{Panopoulou:2019, planck2016-l11B}. We adopt a fiducial value of $\left(p_V/\tau_V\right)_{\rm max} \simeq 0.13\left(2.5\log_{10}e\right)/R_V^{\rm ran} = 0.046$, where here we have approximated $\tau_V\left(\psi = 90^\circ\right) \simeq \tau^{\rm ran}_V$ as the difference between these quantities is numerically small relative to the observational uncertainty on $p_V/E(B-V)$. Since the amplitude of $R_V$ variation with $\psi$ scales approximately linearly with $\left(p_V/\tau_V\right)_{\rm max}$ (Equation~\eqref{eq:rv_psi}), indications that $p_V/E(B-V)$ can exceed 13\%\,mag$^{-1}$ imply that the difference between $R_V\left(0\right)$ and $R_V\left(90^\circ\right)$ computed here is likely a lower limit.

With these values, Equation~\eqref{eq:rv_psi} can be evaluated for all $\psi$, as illustrated in Figure~\ref{fig:rv}. The trend is that $R_V$ becomes smaller---i.e., the extinction law becomes steeper---as $\psi \rightarrow 90^\circ$. This is because $p/\tau$ is larger at $V$ band than at $B$ band while $\tau$ itself is larger at $B$ than $V$. The modulation with $\psi$ makes a relatively small difference to the $B$ band extinction whereas the $V$ band extinction becomes appreciably smaller as $\psi$ increases. Thus, the curve steepens.

Solely from the effect of magnetic field orientation, $R_V$ varies from 3.21 ($\psi = 0$) to 3.05 ($\psi = 90^\circ$). For a random distribution of $\psi$ values (i.e., $\cos\psi$ uniformly distributed between 0 and 1), Equation~\eqref{eq:rv_psi} yields $\sigma\left(R_V\right) = 0.048$. Employing tens of thousands of stellar spectra across the Galactic plane, \citet{Schlafly:2016} found that $R_V$ measurements had a standard deviation $\sigma\left(R_V\right) = 0.18$, comparable to the total range of 0.16 expected from $\psi$ variations. At high Galactic latitudes where $R_V$ is more uniform \citep{Zhang:2024}, it is plausible that $\psi$ is a dominant driver of variation in $R_V$.

Even if the amplitude of this effect has been slightly underestimated in the present work, it is unlikely that $\psi$ variations could account for the entirety of observed $R_V$ variations, especially in the Galactic plane. Most notably, non-uniformity of the magnetic field orientation along the line of sight, an effect attested in data \citep{Clark:2018}, tends to average down the effect of $\psi$ variations. The observed extinction curve must have an $R_V$ intermediate between the highest and lowest $R_V$ values along the line of sight. Indeed, the $R_V$ variation across individual voxels is larger than the variation in $R_V$ as integrated over the line of sight \citep{Zhang:2024}. However, most of the high Galactic latitude sky has only one region that dominates the dust column \citep{Panopoulou:2020}, and so it is here that we predict $\psi$ variations to have the most pronounced effect on $R_V$.

Figure~\ref{fig:ext} presents the $\psi = 0$ and $\psi=90^\circ$ extinction curves normalized to the orientation averaged extinction curve (i.e., $\tau^{\rm ran}$ or equivalently $\sin^2\psi = 2/3$). These are derived from Equation~\eqref{eq:tau_p} using $\tau^{\rm ran}\left(\lambda\right)$ and $p\left(\lambda, 90^\circ\right)$ of the diffuse Galactic ISM as compiled by \citet{Hensley:2021}. In brief, $\tau^{\rm ran}\left(\lambda\right)$ is a synthesis of the extinction curves of \citet{Cardelli:1989}, \citet{Indebetouw:2005}, \citet{Gordon:2009}, \citet{Schlafly:2016}, \citet{Fitzpatrick:2019}, and \citet{Hensley:2020}. $p\left(\lambda\right)$ is based on the recommended Serkowski law of \citet{Whittet:2003} joined to a power law $p\left(\lambda\right) \propto \lambda^{-1.6}$ for $\lambda > 1.38\,\mu$m \citep{Martin:1992}. A $\left(p_V/E(B-V)\right)_{\rm max}$ = 0.13\,mag$^{-1}$ was assumed \citep{Panopoulou:2019, planck2016-l11B}. Given the systematic uncertainties in current determinations of $\left(p_V/E(B-V)\right)_{\rm max}$ driven predominantly by disagreements among reddening maps \citep{Panopoulou:2019}, Figure~\ref{fig:ext} includes a shaded band demonstrating the effect of varying the adopted value of $\left(p_V/E(B-V)\right)_{\rm max}$ between 0.13 and 0.182\,mag$^{-1}$. The latter value corresponds to the best fit of \citet{Panopoulou:2019} to the \citet{schlegel1998} reddening map and the highest value obtained across all maps.

Figure~\ref{fig:ext} indicates that the dependence of $\tau$ on $\psi$ is most pronounced at $\lambda > 1\,\mu$m. The sharp convergence of the $\psi = 0$ and $\psi = 90^\circ$ extinction curves at $\lambda < 1\,\mu$m is driven by the transition of the Serkowski law from increasing to decreasing toward short wavelengths: unaligned grains contribute an increasingly large fraction of the total extinction as $\lambda$ decreases. In contrast, the disparity between the $\psi = 0$ and $\psi = 90^\circ$ curves persists to as long wavelength as $p\left(\lambda\right)$ has been reliably measured. Extinction at these wavelengths is dominated by aligned grains. $R_V$ is defined over a wavelength range where $\tau\left(\lambda, \psi\right)/\tau^{\rm ran}\left(\lambda\right)$ is changing rapidly, yielding the significant change in $R_V$ with $\psi$ presented in Figure~\ref{fig:rv}.

The $R_V$--$\psi$ relation does not impose new constraints on dust models. Any model that can reproduce both $\tau^{\rm ran}\left(\lambda\right)$ and $p\left(\lambda, 90^\circ\right)$ will reproduce the $R_V$--$\psi$ relation automatically by virtue of Equation~\eqref{eq:tau_p}. Nevertheless, Figure~\ref{fig:ext} is a model-independent illustration of the relative contributions of aligned and unaligned grains to interstellar extinction as a function of wavelength.

\section{Connection to Polarized Emission} \label{sec:pol}
Equation~\eqref{eq:rv_psi} and Figure~\ref{fig:rv} imply a nearly linear relationship between $R_V$ and $\sin^2\psi$. Since $\psi$ is not directly observable, in this section we consider other observable quantities sensitive to $\psi$ that can test the $R_V$--$\psi$ relationship developed here. Since polarized emission from dust depends on the orientation of the magnetic field, it is a natural choice.

To develop the theory of polarized emission from a population of partially aligned grains, we again rely on the framework of \citet{Hensley:2019}. Unlike the extinction formalism presented in Section~\ref{sec:theory}, polarized emission is comfortably in the regime $a/\lambda \ll 1$, i.e., the ``electric dipole limit,'' in which the MPFA is asymptotically exact \citep{Draine:2021}. We assume that all dust grains emitting polarized radiation have temperature $T_d$. In current dust models \citep[e.g.,][]{Guillet:2018, Hensley:2023, Ysard:2024}, grains that undergo stochastic heating are too small to be aligned and contribute little to the total emission at far-infrared/submillimeter wavelengths where polarized dust emission is currently best measured. Thus, this is a reasonable approximation. We further assume that grains have size-independent mass density $\rho$. Finally, as with the discussion of extinction, we restrict ourselves to the case of only one emitting region along the line of sight.

It is convenient to define the opacities

\begin{align}
    \kappa^{\rm ran} &= \frac{C_{\rm abs}^{\rm ran}/V}{\rho} \\
    \kappa^{\rm pol} &= \frac{C_{\rm abs}^{\rm pol}/V}{\rho}\,,
\end{align}
where the absorption cross sections $C_{\rm abs}^{\rm ran}$ and $C_{\rm abs}^{\rm pol}$ are defined analogously to the extinction cross sections in Equations~\eqref{eq:cext_ran} and \eqref{eq:cext_pol}, $V$ is the volume of a dust grain, and $\nu$ is frequency. In the electric dipole limit, $C_{\rm abs}/V$ is independent of grain size, and thus so are $\kappa^{\rm ran}$ and $\kappa^{\rm pol}$.

Under these assumptions, a population of dust grains with mass surface density $\Sigma_d$ and temperature $T_d$ emits total and polarized specific intensities $I_\nu$ and $P_\nu$, respectively, at frequency $\nu$

\begin{align}
    I_\nu &= \Sigma_d B_\nu\left(T_d\right)\left[\kappa_\nu^{\rm ran} + f\kappa_\nu^{\rm pol}\left(\frac{2}{3}-\sin^2\psi\right)\right]\\
    P_\nu &= \Sigma_d B_\nu\left(T_d\right) f \kappa_\nu^{\rm pol}\sin^2\psi\,,
\end{align}
where, as in Section~\ref{sec:theory}, $f$ is the fraction of grains that are aligned.

Defining the polarization fraction $\tilde{p}_\nu \equiv P_\nu/I_\nu$, it is straightforward to show for $f > 0$ that for a single dust population

\begin{equation} \label{eq:pfrac}
    \frac{\tilde{p}_\nu}{1+\tilde{p}_\nu} = \frac{\sin^2\psi}{2/3 + \kappa_\nu^{\rm ran}/f \kappa_\nu^{\rm pol}}\,.
\end{equation}

Therefore, if the intrinsic dust properties (i.e., the opacities) are not changing across the sky and if the grain alignment properties (i.e., $f$) are likewise not changing, then $\tilde{p}/(1+\tilde{p})$ is linearly proportional to $\sin^2\psi$. If the total and polarized dust emission arise from a single dust population, then the observed polarization fraction is approximately linearly correlated with $R_V$.

Measurements of the polarization fraction are often complicated by uncertain zero levels, contamination from other sources of emission, and a positive noise bias arising from the fact that $P_\nu$ and $\tilde{p}_\nu$ are positive-definite quantities \citep{planck2014-a12, planck2016-l11B}. Further, Equation~\eqref{eq:pfrac} relies on perfect cancellation of both $\Sigma_d$ and $B_\nu\left(T_d\right)$ between the total and polarized intensities, which may not be true in detail. For instance, the steady-state temperature of large grains has a mild dependence on grain size \citep{Li:2001}.

An alternative observable to the polarization fraction $\tilde{p}_\nu$ is the polarization angle dispersion function $\mathcal{S}$, defined as the standard deviation of the polarization angles observed in a specified region of sky. Because $\mathcal{S}$ depends only on polarization angles, it is completely insensitive to the dust column density or dust temperature. Depending on the region of sky observed, studies have found different power law relationships between $\tilde{p}_\nu$ and $\mathcal{S}$ \citep{Fissel:2016, Hensley:2019, planck2016-l11B}. At high Galactic latitudes where the dust emission is most likely to arise from a single region, \citet{Hensley:2019} found that $\sin^2\psi \propto \mathcal{S}^{-1/2}$. Therefore, we predict that $R_V \propto \mathcal{S}^{-1/2}$ over this sky area.

\section{Summary} \label{sec:summary}

The principal conclusions of this work are as follows:

\begin{enumerate}
    \item It is demonstrated that the wavelength dependence of the interstellar extinction curve, parameterized by $R_V$, varies with the angle $\psi$ between the magnetic field aligning the grains and the line of sight.
    \item The amplitude of this variation is described by a simple parametric $R_V$--$\psi$ relation (Equation~\eqref{eq:rv_psi}) that depends only on quantities that have been well-constrained observationally.
    \item The $R_V$--$\psi$ relation implies that $R_V$ varies between 3.21 and 3.05 as $\psi$ varies from 0 to $90^\circ$ in the diffuse Galactic ISM.
    \item The amplitude of this variation is comparable to the standard deviation of the distribution of $R_V$ values observed over large sky areas from stellar spectroscopy \citep[$\sigma\left(R_V\right) = 0.18$,][]{Schlafly:2016}. This suggests that $\psi$ variation is one of the dominant drivers of $R_V$ variation across the sky, particularly at high Galactic latitudes.
    \item The dust polarization fraction $\tilde{p}_\nu$ and the dust polarization angle dispersion function $\mathcal{S}$ are both sensitive to $\psi$ and thus are both correlated with $R_V$. At high Galactic latitudes, we predict $\tilde{p}_\nu/\left(1+\tilde{p}_\nu\right) \propto R_V$ and $R_V \propto \mathcal{S}^{-1/2}$.
\end{enumerate}

\section*{Acknowledgments}
It is a pleasure to thank Xiangyu Zhang, Greg Green, and Bruce Draine for stimulating discussions and helpful feedback, and the anonymous referee for suggestions that improved this work. A special thanks to Vincent Pelgrims for pointing out the work of Nikolai Voshchinnikov and collaborators on this topic. The research was carried out at the Jet Propulsion Laboratory, California Institute of Technology, under a contract with the National Aeronautics and Space Administration (80NM0018D0004).

\software{Matplotlib \citep{Matplotlib}, NumPy \citep{NumPy}}

\bibliography{refs, Planck_bib}

\begin{thebibliography}{}
\expandafter\ifx\csname natexlab\endcsname\relax\def\natexlab#1{#1}\fi
\providecommand{\url}[1]{\href{#1}{#1}}
\providecommand{\dodoi}[1]{doi:~\href{http://doi.org/#1}{\nolinkurl{#1}}}
\providecommand{\doeprint}[1]{\href{http://ascl.net/#1}{\nolinkurl{http://ascl.net/#1}}}
\providecommand{\doarXiv}[1]{\href{https://arxiv.org/abs/#1}{\nolinkurl{https://arxiv.org/abs/#1}}}

\bibitem[{{Cardelli} {et~al.}(1989){Cardelli}, {Clayton}, \& {Mathis}}]{Cardelli:1989}
{Cardelli}, J.~A., {Clayton}, G.~C., \& {Mathis}, J.~S. 1989, \apj, 345, 245, \dodoi{10.1086/167900}

\bibitem[{{Clark}(2018)}]{Clark:2018}
{Clark}, S.~E. 2018, \apjl, 857, L10, \dodoi{10.3847/2041-8213/aabb54}

\bibitem[{{Clayton} {et~al.}(1992){Clayton}, {Anderson}, {Magalhaes}, {Code}, {Nordsieck}, {Meade}, {Wolff}, {Babler}, {Bjorkman}, {Schulte-Ladbeck}, {Taylor}, \& {Whitney}}]{Clayton:1992}
{Clayton}, G.~C., {Anderson}, C.~M., {Magalhaes}, A.~M., {et~al.} 1992, \apjl, 385, L53, \dodoi{10.1086/186276}

\bibitem[{{Dolginov} \& {Mitrofanov}(1976)}]{Dolginov:1976}
{Dolginov}, A.~Z., \& {Mitrofanov}, I.~G. 1976, \apss, 43, 291, \dodoi{10.1007/BF00640010}

\bibitem[{{Draine} \& {Hensley}(2021)}]{Draine:2021}
{Draine}, B.~T., \& {Hensley}, B.~S. 2021, \apj, 919, 65, \dodoi{10.3847/1538-4357/ac0050}

\bibitem[{{Draine} \& {Weingartner}(1996)}]{Draine:1996}
{Draine}, B.~T., \& {Weingartner}, J.~C. 1996, \apj, 470, 551, \dodoi{10.1086/177887}

\bibitem[{{Draine} \& {Weingartner}(1997)}]{Draine:1997}
---. 1997, \apj, 480, 633, \dodoi{10.1086/304008}

\bibitem[{{Fissel} {et~al.}(2016){Fissel}, {Ade}, {Angil{\`e}}, {Ashton}, {Benton}, {Devlin}, {Dober}, {Fukui}, {Galitzki}, {Gandilo}, {Klein}, {Korotkov}, {Li}, {Martin}, {Matthews}, {Moncelsi}, {Nakamura}, {Netterfield}, {Novak}, {Pascale}, {Poidevin}, {Santos}, {Savini}, {Scott}, {Shariff}, {Diego Soler}, {Thomas}, {Tucker}, {Tucker}, \& {Ward-Thompson}}]{Fissel:2016}
{Fissel}, L.~M., {Ade}, P. A.~R., {Angil{\`e}}, F.~E., {et~al.} 2016, \apj, 824, 134, \dodoi{10.3847/0004-637X/824/2/134}

\bibitem[{{Fitzpatrick} {et~al.}(2019){Fitzpatrick}, {Massa}, {Gordon}, {Bohlin}, \& {Clayton}}]{Fitzpatrick:2019}
{Fitzpatrick}, E.~L., {Massa}, D., {Gordon}, K.~D., {Bohlin}, R., \& {Clayton}, G.~C. 2019, \apj, 886, 108, \dodoi{10.3847/1538-4357/ab4c3a}

\bibitem[{{Gordon} {et~al.}(2009){Gordon}, {Cartledge}, \& {Clayton}}]{Gordon:2009}
{Gordon}, K.~D., {Cartledge}, S., \& {Clayton}, G.~C. 2009, \apj, 705, 1320, \dodoi{10.1088/0004-637X/705/2/1320}

\bibitem[{{Greenberg}(1960)}]{Greenberg:1960}
{Greenberg}, J.~M. 1960, Journal of Applied Physics, 31, 82, \dodoi{10.1063/1.1735423}

\bibitem[{{Greenberg}(1968)}]{Greenberg:1968}
---. 1968, in Nebulae and Interstellar Matter, ed. B.~M. {Middlehurst} \& L.~H. {Aller}, 221

\bibitem[{{Greenberg} \& {Chlewicki}(1987)}]{Greenberg:1987}
{Greenberg}, J.~M., \& {Chlewicki}, G. 1987, \qjras, 28, 321

\bibitem[{{Greenberg} \& {Meltzer}(1960)}]{Greenberg:1960b}
{Greenberg}, J.~M., \& {Meltzer}, A.~S. 1960, \apj, 132, 667, \dodoi{10.1086/146970}

\bibitem[{{Guillet} {et~al.}(2018){Guillet}, {Fanciullo}, {Verstraete}, {Boulanger}, {Jones}, {Miville-Desch{\^e}nes}, {Ysard}, {Levrier}, \& {Alves}}]{Guillet:2018}
{Guillet}, V., {Fanciullo}, L., {Verstraete}, L., {et~al.} 2018, \aap, 610, A16, \dodoi{10.1051/0004-6361/201630271}

\bibitem[{{Hall}(1949)}]{Hall:1949}
{Hall}, J.~S. 1949, Science, 109, 166, \dodoi{10.1126/science.109.2825.166}

\bibitem[{{Hensley} \& {Draine}(2020)}]{Hensley:2020}
{Hensley}, B.~S., \& {Draine}, B.~T. 2020, \apj, 895, 38, \dodoi{10.3847/1538-4357/ab8cc3}

\bibitem[{{Hensley} \& {Draine}(2021)}]{Hensley:2021}
---. 2021, \apj, 906, 73, \dodoi{10.3847/1538-4357/abc8f1}

\bibitem[{{Hensley} \& {Draine}(2023)}]{Hensley:2023}
---. 2023, \apj, 948, 55, \dodoi{10.3847/1538-4357/acc4c2}

\bibitem[{{Hensley} {et~al.}(2019){Hensley}, {Zhang}, \& {Bock}}]{Hensley:2019}
{Hensley}, B.~S., {Zhang}, C., \& {Bock}, J.~J. 2019, \apj, 887, 159, \dodoi{10.3847/1538-4357/ab5183}

\bibitem[{{Hiltner}(1949)}]{Hiltner:1949}
{Hiltner}, W.~A. 1949, Science, 109, 165, \dodoi{10.1126/science.109.2825.165}

\bibitem[{{Hoang} \& {Lazarian}(2016)}]{Hoang:2016}
{Hoang}, T., \& {Lazarian}, A. 2016, \apj, 831, 159, \dodoi{10.3847/0004-637X/831/2/159}

\bibitem[{{Hunter}(2007)}]{Matplotlib}
{Hunter}, J.~D. 2007, Computing in Science and Engineering, 9, 90, \dodoi{10.1109/MCSE.2007.55}

\bibitem[{{Indebetouw} {et~al.}(2005){Indebetouw}, {Mathis}, {Babler}, {Meade}, {Watson}, {Whitney}, {Wolff}, {Wolfire}, {Cohen}, {Bania}, {Benjamin}, {Clemens}, {Dickey}, {Jackson}, {Kobulnicky}, {Marston}, {Mercer}, {Stauffer}, {Stolovy}, \& {Churchwell}}]{Indebetouw:2005}
{Indebetouw}, R., {Mathis}, J.~S., {Babler}, B.~L., {et~al.} 2005, \apj, 619, 931, \dodoi{10.1086/426679}

\bibitem[{{Karim} {et~al.}(2024){Karim}, {Singh}, {Rezaie}, {Eisenstein}, {Hadzhiyska}, {Speagle}, {Aguilar}, {Ahlen}, {Brooks}, {Claybaugh}, {de la Macorra}, {Ferraro}, {Forero-Romero}, {Gazta{\~n}aga}, {Gontcho}, {Gutierrez}, {Guy}, {Honscheid}, {Juneau}, {Kirkby}, {Krolewski}, {Lambert}, {Landriau}, {Levi}, {Meisner}, {Miquel}, {Moustakas}, {Mu{\~n}oz-Guti{\'e}rrez}, {Myers}, {Niz}, {Palanque Delabrouille}, {Percival}, {Prada}, {Rossi}, {Sanchez}, {Schlafly}, {Schlegel}, {Schubnell}, {Sprayberry}, {Tarl{\'e}}, {Weaver}, \& {Zou}}]{Karim:2024}
{Karim}, T., {Singh}, S., {Rezaie}, M., {et~al.} 2024, arXiv e-prints, arXiv:2408.15909, \dodoi{10.48550/arXiv.2408.15909}

\bibitem[{{Kim} \& {Martin}(1995)}]{Kim:1995}
{Kim}, S.-H., \& {Martin}, P.~G. 1995, \apj, 444, 293, \dodoi{10.1086/175604}

\bibitem[{{Lee} \& {Draine}(1985)}]{Lee:1985}
{Lee}, H.~M., \& {Draine}, B.~T. 1985, \apj, 290, 211, \dodoi{10.1086/162974}

\bibitem[{{Li} \& {Draine}(2001)}]{Li:2001}
{Li}, A., \& {Draine}, B.~T. 2001, \apj, 554, 778, \dodoi{10.1086/323147}

\bibitem[{{Lillie} \& {Witt}(1976)}]{Lillie:1976}
{Lillie}, C.~F., \& {Witt}, A.~N. 1976, \apj, 208, 64, \dodoi{10.1086/154582}

\bibitem[{{Martin} {et~al.}(1999){Martin}, {Clayton}, \& {Wolff}}]{Martin:1999}
{Martin}, P.~G., {Clayton}, G.~C., \& {Wolff}, M.~J. 1999, \apj, 510, 905, \dodoi{10.1086/306613}

\bibitem[{{Martin} {et~al.}(1992){Martin}, {Adamson}, {Whittet}, {Hough}, {Bailey}, {Kim}, {Sato}, {Tamura}, \& {Yamashita}}]{Martin:1992}
{Martin}, P.~G., {Adamson}, A.~J., {Whittet}, D.~C.~B., {et~al.} 1992, \apj, 392, 691, \dodoi{10.1086/171470}

\bibitem[{{Panopoulou} {et~al.}(2019){Panopoulou}, {Hensley}, {Skalidis}, {Blinov}, \& {Tassis}}]{Panopoulou:2019}
{Panopoulou}, G.~V., {Hensley}, B.~S., {Skalidis}, R., {Blinov}, D., \& {Tassis}, K. 2019, \aap, 624, L8, \dodoi{10.1051/0004-6361/201935266}

\bibitem[{{Panopoulou} \& {Lenz}(2020)}]{Panopoulou:2020}
{Panopoulou}, G.~V., \& {Lenz}, D. 2020, \apj, 902, 120, \dodoi{10.3847/1538-4357/abb6f5}

\bibitem[{{\sorthelp{Planck Collaboration 2015J}}{Planck Collaboration X}(2016)}]{planck2014-a12}
{\sorthelp{Planck Collaboration 2015J}}{Planck Collaboration X}. 2016, \aap, 594, A10, \dodoi{10.1051/0004-6361/201525967}

\bibitem[{{\sorthelp{Planck Collaboration 2018L}}{Planck Collaboration XII}(2020)}]{planck2016-l11B}
{\sorthelp{Planck Collaboration 2018L}}{Planck Collaboration XII}. 2020, \aap, 641, A12, \dodoi{10.1051/0004-6361/201833885}

\bibitem[{{\sorthelp{Planck Collaboration IntS}}{Planck Collaboration Int. XIX}(2015)}]{planck2014-XIX}
{\sorthelp{Planck Collaboration IntS}}{Planck Collaboration Int. XIX}. 2015, \aap, 576, A104, \dodoi{10.1051/0004-6361/201424082}

\bibitem[{{Purcell}(1975)}]{Purcell:1975}
{Purcell}, E.~M. 1975, in The Dusty Universe, ed. G.~B. {Field} \& A.~G.~W. {Cameron}, 155--167

\bibitem[{{Purcell}(1979)}]{Purcell:1979}
---. 1979, \apj, 231, 404, \dodoi{10.1086/157204}

\bibitem[{{Schlafly} {et~al.}(2016){Schlafly}, {Meisner}, {Stutz}, {Kainulainen}, {Peek}, {Tchernyshyov}, {Rix}, {Finkbeiner}, {Covey}, {Green}, {Bell}, {Burgett}, {Chambers}, {Draper}, {Flewelling}, {Hodapp}, {Kaiser}, {Magnier}, {Martin}, {Metcalfe}, {Wainscoat}, \& {Waters}}]{Schlafly:2016}
{Schlafly}, E.~F., {Meisner}, A.~M., {Stutz}, A.~M., {et~al.} 2016, \apj, 821, 78, \dodoi{10.3847/0004-637X/821/2/78}

\bibitem[{{Schlegel} {et~al.}(1998){Schlegel}, {Finkbeiner}, \& {Davis}}]{schlegel1998}
{Schlegel}, D.~J., {Finkbeiner}, D.~P., \& {Davis}, M. 1998, \apj, 500, 525, \dodoi{10.1086/305772}

\bibitem[{{Serkowski} {et~al.}(1975){Serkowski}, {Mathewson}, \& {Ford}}]{Serkowski:1975}
{Serkowski}, K., {Mathewson}, D.~S., \& {Ford}, V.~L. 1975, \apj, 196, 261, \dodoi{10.1086/153410}

\bibitem[{{Siebenmorgen}(2023)}]{Siebenmorgen:2023}
{Siebenmorgen}, R. 2023, \aap, 670, A115, \dodoi{10.1051/0004-6361/202243860}

\bibitem[{{Trumpler}(1930)}]{Trumpler:1930}
{Trumpler}, R.~J. 1930, Lick Observatory Bulletin, 420, 154, \dodoi{10.5479/ADS/bib/1930LicOB.14.154T}

\bibitem[{{van der Walt} {et~al.}(2011){van der Walt}, {Colbert}, \& {Varoquaux}}]{NumPy}
{van der Walt}, S., {Colbert}, S.~C., \& {Varoquaux}, G. 2011, Computing in Science and Engineering, 13, 22, \dodoi{10.1109/MCSE.2011.37}

\bibitem[{{Voshchinnikov}(1989)}]{Voshchinnikov:1989}
{Voshchinnikov}, N.~V. 1989, Astronomische Nachrichten, 310, 265, \dodoi{10.1002/asna.2113100405}

\bibitem[{{Weingartner} \& {Draine}(2001)}]{Weingartner:2001}
{Weingartner}, J.~C., \& {Draine}, B.~T. 2001, \apj, 548, 296, \dodoi{10.1086/318651}

\bibitem[{{Whittet}(2003)}]{Whittet:2003}
{Whittet}, D.~C.~B., ed. 2003, {Dust in the galactic environment}

\bibitem[{{Wolff} {et~al.}(1997){Wolff}, {Clayton}, {Kim}, {Martin}, \& {Anderson}}]{Wolff:1997}
{Wolff}, M.~J., {Clayton}, G.~C., {Kim}, S.-H., {Martin}, P.~G., \& {Anderson}, C.~M. 1997, \apj, 478, 395, \dodoi{10.1086/303789}

\bibitem[{{Ysard} {et~al.}(2024){Ysard}, {Jones}, {Guillet}, {Demyk}, {Decleir}, {Verstraete}, {Choubani}, {Miville-Desch{\^e}nes}, \& {Fanciullo}}]{Ysard:2024}
{Ysard}, N., {Jones}, A.~P., {Guillet}, V., {et~al.} 2024, \aap, 684, A34, \dodoi{10.1051/0004-6361/202348391}

\bibitem[{{Zhang} \& {Green}(2024)}]{Zhang:2024}
{Zhang}, X., \& {Green}, G. 2024, arXiv e-prints, arXiv:2407.14594, \dodoi{10.48550/arXiv.2407.14594}

\end{thebibliography}

\end{document}